\documentclass[prl, superscriptaddress, twocolumn, showpacs]{revtex4-1}

\usepackage{amsfonts}
\usepackage{amssymb}
\usepackage[dvips]{graphicx}
\usepackage{amsmath}
\usepackage{pstricks}

\newcommand{\bra}[1] {\langle #1|}
\newcommand{\ket}[1] {| #1 \rangle}

\begin{document}

\title{Real-world Quantum Sensors: Evaluating Resources for Precision Measurement}

\author{N.~Thomas-Peter}
\affiliation{Clarendon Laboratory, University of Oxford, Parks Road, Oxford OX1 3PU, United Kingdom}
\author{B.~J.~Smith} 
\affiliation{Clarendon Laboratory, University of Oxford, Parks Road, Oxford OX1 3PU, United Kingdom}
\author{A.~Datta}
\affiliation{Clarendon Laboratory, University of Oxford, Parks Road, Oxford OX1 3PU, United Kingdom}
\author{L.~Zhang}
\affiliation{Clarendon Laboratory, University of Oxford, Parks Road, Oxford OX1 3PU, United Kingdom}
\author{U.~Dorner}
\affiliation{Centre for Quantum Technologies, National University of Singapore, 3 Science Drive 2, 117543 Singapore, Singapore}
\affiliation{Clarendon Laboratory, University of Oxford, Parks Road, Oxford OX1 3PU, United Kingdom}
\author{I.~A.~Walmsley}
\affiliation{Clarendon Laboratory, University of Oxford, Parks Road, Oxford OX1 3PU, United Kingdom}


\begin{abstract}
Quantum physics holds the promise of enabling certain tasks with better performance than possible when only classical resources are employed.  The quantum phenomena present in many experiments signify nonclassical behavior, but do not always imply superior performance.  Quantifying the enhancement  achieved from quantum behavior requires careful analysis of the resources involved. We analyze the specific case of parameter estimation using an optical interferometer, where increased precision can be achieved using quantum probe states. Common performance measures are examined and it is shown that some overestimate the improvement. For the simplest experimental case we compare the different measures and show this overestimate explicitly. We give the preferred analysis of real-world experiments and calculate benchmark values for experimental parameters necessary to realize a precision enhancement.

\end{abstract}

\maketitle

Quantum-enhanced technologies such as quantum computing, cryptography and metrology utilize nonclassical behavior of quantum systems to surpass the performance of their classical analogs. Often signatures of quantum behavior such as violation of a Bell inequality, photon antibunching, or sub-Poissonian number statistics are observed in an experiment.  In some cases, their presence may be sufficient to ensure superior performance of the quantum protocol over the classical.  There are situations, however, in which quantum behavior does not necessarily imply performance beyond classical limits and further analysis is necessary to ensure improvement.

An important class of quantum-enhanced technologies involves sensing, in which quantum probes are used to estimate a parameter of a physical system with precision beyond what is possible with classical resources~\cite{glm04}. Analysis of the limits on precision, something which is of both practical and fundamental interest, requires careful accounting of the necessary resources to compare classical and quantum protocols. Entangled states of multiple particles are a critical component in enabling quantum enhancement in many situations~\cite{glm06}, as are detectors that extract maximum information about the parameter of interest~\cite{bc94}, what we call optimal measurements.  Calculating the precision that can be achieved for fixed resources provides a basis for comparing quantum and classical protocols.  Recent work has shown that practical imperfections such as loss, decoherence, state-preparation and detector inefficiency decrease the enhancement of quantum metrology protocols \cite{dorner:09,datta:10, Huelga97, *Shaji07, *Rubin07, *Huver08, *Gilbert08}.  Therefore poor enumeration, which distorts the resources used and hence the enhancement obtained, is problematic.  

Several proof-of-principle experiments aimed at quantum-enhanced parameter estimation have been performed to date~\cite{mls04, *higgins:07, *eisenberg:05, *resch:07-1, *nagata:07, *smith:08, *smith:09, *mpso09, *kim:09, *matthews:10}. Many utilize nonclassical signatures in detection outcomes to indicate improved precision. However, as we will show, this is insufficient to demonstrate increased precision beyond the classical limit and further information must be included to accurately compare quantum and classical strategies.

In this Letter, we examine a quantum sensor based on optical interferometry. We evaluate the resources, in terms of number of trials and photons, to reach a given precision when estimating the phase between two optical modes. Limits to this precision are analysed using Fisher information and quantum Fisher information and compared with other common measures of enhancement such as super-sensitivity~\cite{resch:07}. When experimental imperfections are present, we show that quantum enhancement can be overestimated by improper resource accounting.  In particular, a common error is to neglect the channel transmission, source, and detection efficiencies. To experimentally demonstrate this, we construct the simplest quantum states which ideally lead to improved precision and examine the behavior of different enhancement measures. We show that improvement in state preparation and detection efficiencies are necessary to unambiguously demonstrate quantum enhancement in phase estimation, and set quantitative benchmarks for these.

To estimate the phase $\phi$ in a two-mode interferometer, a state $\hat{\rho}$ is launched into it, evolves into the state $\hat\rho(\phi)$ due to interaction with the phase-shifting element, and a measurement is performed at the sensor output. The phase $\phi$ is estimated from the outcomes of $\nu$ repeated experimental trials. We define the average number of photons at the interferometer \emph{input}, $N=\langle \hat{n} \rangle$, and the total number of experimental trials, $\nu$, necessary to achieve a given level of phase precision, $\Delta \phi$, as the resources.  For fixed resources, we then compare the precision of the quantum and classical approaches and the approach with the smallest phase uncertainty is thus the better strategy.

Two effects attributed to quantum behavior in interferometry are phase super-resolution and phase super-sensitivity. Phase super-resolution, the sinusoidal variation of an $N$-fold detection signal as a function of interferometer phase $\phi$ with an $N$-times increase in rate of oscillation, can be observed using only classical input states of light and projective measurements \cite{resch:07}. It has recently been shown that the visibility of such super-resolving fringes can be used to distinguish between quantum and classical input states of a Mach-Zehnder interferometer \cite{afek:10-1}. This signature of quantum behavior does not, however, quantify improved performance beyond classical interferometry.

Phase super-sensitivity, a commonly employed measure of performance, is defined as reduced phase uncertainty compared to that possible with classical resources~\cite{resch:07}.  The model employed in Ref.~\cite{resch:07} incorporates experimental imperfection through `efficiency' and visibility parameters $\eta_{\rm{p}}$ and $V$,  by means of a phenomenological model of an $N_{\rm{d}}$-fold detected coincidence signal. Here, $\eta_{\rm{p}}$ is the proportion of the input state $\hat{\rho}$ that can lead to an $N_{\rm{d}}$-fold detection event.  The visibility is required to satisfy $\eta_\mathrm{p}V^{2}N_\mathrm{d} > 1$ in order for the measurement to be regarded as `super-sensitive'.  However, this assumes unit sensor transmission and perfectly efficient detectors.  It includes only imperfect state preparation; with $N_\mathrm{d} = N$, the number of input photons.  In fact, the most common experimental situation is that $N_\mathrm{d} < N$ so that the resources consumed are significantly underestimated.  Claims of precision beyond the classical limit that utilize this measure must therefore be interpreted carefully.

In general, the precision of the phase estimate is limited by the Cram\'er-Rao bound (CRB)~\cite{bc94}, $\Delta \phi\ge 1/\sqrt{\nu F(\phi)}$, where  $F(\phi) = \sum_j (1/p_j( \phi )) \left|\partial p_j (\phi)/\partial \phi\right|^{2}$, is the Fisher information (FI). The probability $p_j(\phi)$ corresponds to an outcome $j$ of a measurement.  $F(\phi)$ is bounded from above by the quantum Fisher information (QFI), $F_Q$, found by maximizing $F(\phi)$ over all physical measurements. Hence the quantum Cram\'er-Rao bound (QCRB), $\Delta \phi \ge 1/\sqrt{\nu F_Q(\phi)}$, depends only on the input state and channel, independent of detector configuration~\cite{bc94}.

In the absence of imperfections, the QCRB of $N$ uncorrelated photons leads to the standard quantum limit $\Delta\phi \ge \Delta\phi_{\mathrm{SQL}} =1/\sqrt{\nu N}$, equivalent to the precision attained with a coherent state of unknown phase with average photon number $|\alpha|^2 = N$. When quantum input states are employed, the best attainable precision is given by the so-called Heisenberg limit $\Delta \phi \ge \Delta \phi_{\mathrm{HL}} = 1/\sqrt{\nu}N$ which is achieved by the widely studied `$N00N$ state'~\cite{resch:07, afek:10-1, sanders:95, *boto:00, *bollinger:96, mls04, *eisenberg:05, *higgins:07, *resch:07-1, *nagata:07, *smith:08, *smith:09, *mpso09, *kim:09, *matthews:10}. Note that to reach this precision, an appropriate optimal detection scheme with unit efficiency must be used. However, if the sensor has finite transmissivity $\eta$, assumed equal in both interferometer arms, the classical strategy is bounded by the standard interferometric limit (SIL) $\Delta\phi \ge \Delta\phi_{\mathrm{SIL}} = 1/\sqrt{\nu \eta N}$~\cite{dorner:09}. $N00N$ states are extremely susceptible to loss; removal of even one photon results in a phase-insensitive state. The QFI for $N00N$ states through the same channel is thus scaled by the probability that all photons are transmitted, $\eta^{N}$, giving a QCRB of $\Delta\phi \ge 1/\sqrt{\nu \eta^{N}}N$. 

For $N00N$ states, an optimal measurement is projection onto $\ket{\Psi_{\pm}} = (|N,0\rangle \pm |0,N\rangle ) / \sqrt{2}$, where $\ket{m,n}$ denotes $m$ ($n$) photons in each mode of the interferometer. This measurement set has three possible outcomes: $j=\pm$ (detection of $\ket{\Psi_\pm}$) and $j=0$ (otherwise). In practice, it is difficult to realize single-mode sensors and this may lead to non-ideal interference due to unmeasured distinguishing information.  Measurement outcomes therefore typically have probabilities $p_{\pm}(\phi) = f[1\pm V\cos(N\phi)]/2$ and $p_{0}(\phi) = 1-p_{+}(\phi) - p_{-}(\phi)$, leading to a CRB $\Delta\phi\ge\Delta\phi_{\mathrm{min}} = 1/\sqrt{\nu f}NV$ \cite{epaps}. Here, $f =\eta_\mathrm{p}(\eta\eta_\mathrm{d})^{N}$, where $\eta_\mathrm{d}$ is detector efficiency, and $V$ is the interference visibility accounting for multimode states and detectors.  The state is said to exhibit super-resolution if $p_{\pm}(\phi)$ oscillates $N$ times the applied phase. To surpass the SIL requires $\Delta \phi_\mathrm{min} < \Delta \phi_\mathrm{SIL}$, leading to a threshold visibility 
\begin{equation}
V_\mathrm{th}=\sqrt{\eta/fN},
\label{eq:1}
\end{equation}
which is equivalent to that in Ref. \cite{resch:07} with $\eta = \eta_\mathrm{d} = 1$ and $N=N_\mathrm{d}$. A state is said to exhibit super-sensitivity if $V\ge V_\mathrm{th}$, and the classical precision limit is surpassed when this occurs. 

The FI and QFI analyses give a straightforward method to quantify the resources consumed in an experiment and compare quantum with classical strategies. The QFI reveals whether or not the classical limit could be beaten in principle using a particular state and channel, while the FI reveals whether or not it can be beaten with the addition of a particular measurement.  To ascertain the best performance of a given quantum strategy, one must therefore know both the density matrix $\hat{\rho}$ of the input state, giving $N$, FI, and QFI, and when it is prepared, giving $\nu$, as well as the device transmission and detector efficiency.  In our case, $\hat\rho$ will be obtained through state tomography and $\nu$ by heralded state preparation. 

A common practice is to `post-select' on particular measurement outcomes and neglect the occurence of others, including when nothing is detected at the output.  This is equivalent to setting $\eta_\mathrm{p}$, $\eta$, $\eta_\mathrm{d}$ to 1 and $N = N_\mathrm{d}$.  This neglects both $\nu$ and the true $N$, significantly underestimating both the number of trials (which grows exponentially in $N$~\cite{epaps}) and information from additional measurement outcomes.

To experimentally examine realistic quantum sensors, we implement the simplest case, employing two-particle states, in which increased precision can be shown. We use an optical sensor design based on a heralded two-photon Holland-Burnett (HB) state \cite{Holland:93, epaps}. HB($N = 2k$) states are prepared by launching $k$ photons in both input ports of a 50:50 beam splitter.  The HB($2$) state provides the least stringent constraints on sensor transmissivity and detection efficiency \cite{dorner:09, datta:10}. Furthermore, the optimal measurement for phase estimation using HB states can be implemented with realistic number-resolving detection \cite{datta:10}. The QCRB of a HB state in the lossless case is $\Delta\phi \ge 1/\sqrt{\nu N(N/2+1)}$ which, for large $N$, has Heisenberg scaling that differs only by a constant factor. With loss this state retains phase sensitivity and precision enhancement far better than $N00N$ states \cite{dunningham:02}. Indeed, the QCRB for the HB state approximately follows the optimum value for a wide range of $\eta$, $\eta_\mathrm{d}$ and $N$~\cite{dorner:09, datta:10}.

\begin{figure}[ht] 
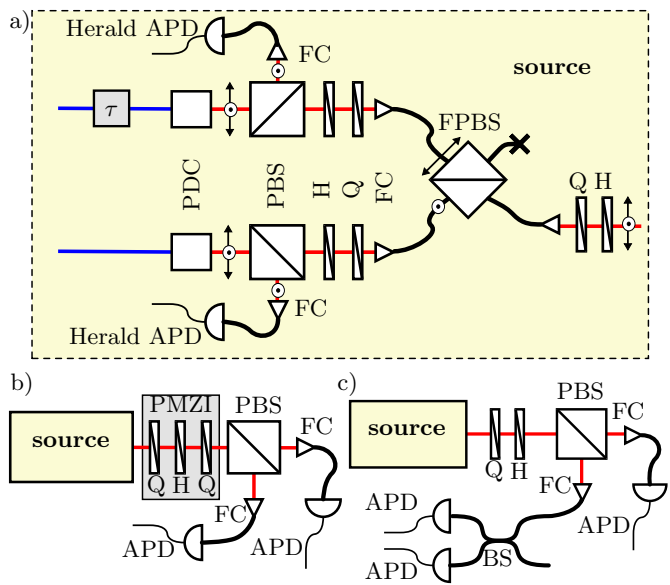
\vspace{-0.7cm}
\caption{
(color online) Setup for generation and characterization of heralded HB states. (a) State generation based on two parametric downconverters (PDC). Half- (H) and a quarter-wave plates (Q) after each polarizing beam splitter (PBS) adjust the polarizations to combine photons at a fiber polarizing beam splitter (FPBS). Coincidence detection between heralding avalanche photodiodes (Herald APDs) signals HB state preparation at the output of the FPBS. (b) PMZI applies phase $\phi$ between $\pm45^{\circ}$ polarizations. Coincidence detection between the APDs implement optimal measurement for HB($2$) state. (c) Tomographic measurements are implemented using a quarter- (Q) and half-wave plate (H) followed by a PBS. Outputs are coupled into single-mode fiber (FC) with the reflected mode split by a 50:50 fiber beam splitter (FBS) to allow partial number resolution.}
 \label{fig:sourceSetup} 
 \vspace{-0.2cm}
\end{figure}

The scheme used to generate heralded HB states is shown in
Fig.~\ref{fig:sourceSetup}(a). A polarization-based $N$-photon HB state
is obtained by combining two orthogonally polarized $k = N/2$-photon Fock
states at a polarizing beam splitter~\cite{epaps}. The required Fock states are
generated by two parametric downconverters (PDC), where $k$ photons
in one mode of each PDC are heralded by detection of $k$ photons in
the other mode. The PDCs are designed to produce spectrally
disentangled modes~\cite{mosley:08} so that heralding does not
compromise the state purity, which is crucial for high-visibility
interferometry. Since spectral filtering is not required to increase
the state purity, the preparation efficiency is determined solely by
the system loss. In principle this source, with the use of
photon-number-resolving heralding detectors~\cite{achilles:03},
generates ideal HB states of arbitrary $N$. 

\begin{figure}[ht] 
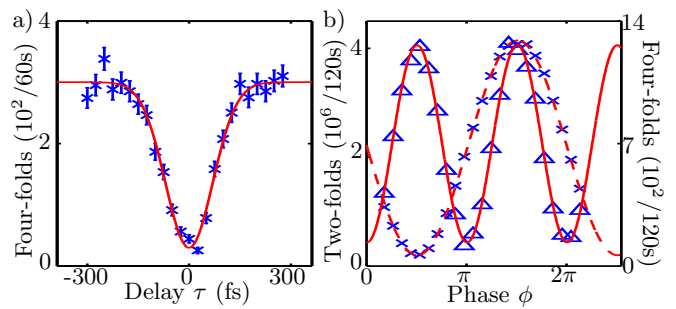
\vspace{-0.4cm}
 \caption{%
 (color online) Quantum intereference of heralded photonic states. (a) Hong-Ou-Mandel interference between two heralded single photons as a function of delay, $\tau$. (b) Setting $\tau=0$ and scanning the PMZI phase shows two-photon (solid) and single-photon (dashed) interference. Error bars are derived from Poissonian statistics and contributions from multiple PDC pair emissions are removed.}
 \vspace{-0.3cm}
 \label{fig:interference}
\end{figure}

The quality of the heralded Fock states used to generate the HB($2$) state was tested by means of Hong-Ou-Mandel interference between heralded single photons (Fig. \ref{fig:interference}a). The theoretical fit gives a visibility of $90\pm3\%$ ($80\pm3\%$ raw), setting a lower bound on the input photon purity and distinguishability, and an upper bound for the multi-photon fringe visibility in Eq.~(\ref{eq:1}). The residual impurity is partly intrinsic~\cite{mosley:08} and partly due to imperfect compensation of the optical fiber birefringence.

\begin{figure}[ht] 
\vspace{-1.2cm}
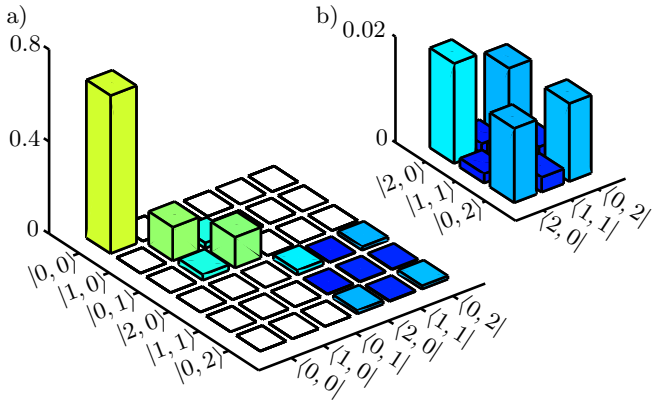
 \caption{%
 (color online) Reconstructed heralded quantum state. (a) The absolute value of the density matrix including vacuum, one-, and two-photon
 subspaces.  (b) An enlargement of the two-photon subspace. In all density matrices, $\ket{m,n}$ denotes
 $m$ diagonally ($+45^{\circ}$) polarized photons and $n$
 anti-diagonally ($-45^{\circ}$) polarized photons. Bars are colored
 according to their magnitude.}
 \label{fig:rho}
 \vspace{-0.5cm}
\end{figure}

We reconstruct the full heralded state by performing state tomography using the experimental arrangement shown in Fig.~\ref{fig:sourceSetup}(c)~\cite{thomas-peter:09, epaps}. Population in lower photon number subspaces arises from loss so that coherences between subspaces of different photon number can be neglected.  Figure~\ref{fig:rho}(a) shows the reconstructed density matrix.  The heralded state has populations of $0.686$, $0.277$, and $0.037$ in the zero-, one-, and two-photon subspaces giving an average photon number of $0.35$.  This state has an overlap of 0.031 with the ideal 2-photon HB state, while the re-normalized two photon subspace has an overlap of 0.85. 

The heralded HB states are launched into a polarization Mach-Zehnder interferometer, PMZI in Fig.~\ref{fig:sourceSetup}(b), which introduces a controllable relative phase between the two modes. Figure \ref{fig:interference}(b) shows the detection count-rate phase dependence for heralded single-photon states (dashed) and heralded HB($2$) states (solid). Theoretical fits show visibilities of $90.9\pm0.5\%$ and $80.6\pm1.5\%$ ($76.4\pm1.4\%$ raw) respectively. The latter is consistent with the measured state purity.

For $N=2$, HB and $N00N$ states are equivalent and Eq.~(\ref{eq:1}) can be applied. If we use a post-selection analysis, only data resulting in $2$-fold outcomes are recorded, $\eta$ and $f$ are set to $1$ and $N=2$. This yields a threshold visibility $V_\mathrm{th} = 1/\sqrt{2}$.  The measured fringe visibility significantly exceeds this bound and so by this analysis we achieve super-sensitivity, beating the SQL.   If we renormalise the post-selected photon number subspace and use $\eta_\mathrm{p}=0.85$ while still neglecting $\eta$ and $\eta_\mathrm{d}$, we obtain $V_\mathrm{th} = 0.77$.  Our measured fringe visibility also exceeds this threshold, again demonstrating super-sensitivity and beating the SQL.  Since the whole state including lower photon number subspaces is known however, the correct $\eta_\mathrm{p}$, given by the overlap of the reconstructed state with the HB(2) state, can be calculated as $\eta_\mathrm{p} = 0.031$.  We can therefore apply the bound in Ref.~\cite{resch:07}, leading to $V_\mathrm{th} = 1/\sqrt{\eta_\mathrm{p}N_\mathrm{d}} = 4.02$.  This is an unphysical visibility and so, by this analysis, we cannot beat the SQL.  The above threshold visibilities are based on post-selection and renormalisation or neglect both channel transmissivity and detector efficiencies.  They therefore significantly underestimate the resources required to reach a given precision.  Heralding a characterised state affords a complete reckoning of the resources used by counting and all states input into the interferometer.  $f$ and $\eta$ can then be estimated experimentally, allowing $V_\mathrm{th}$ to be calculated according to Eq.~(\ref{eq:1}).  $f$ is determined from the ratio of four-fold coincidences to heralding events, giving $f=0.0047\pm0.0001$ and $\eta\eta_\mathrm{d} = \sqrt{f/\eta_\mathrm{p}} = 0.39$.  For our detector efficiencies of $\eta_p = 0.45\%$, $\eta = 0.87$ which results in $V_\mathrm{th} = \sqrt{\eta/Nf} = 9.57$, an even higher threshold on visibility, showing that we cannot beat the SIL.  This explicitly shows that to unequivocally demonstrate performance beyond the classical limit, all system inefficiencies must be characterised.  

Equation~(\ref{eq:1}) and the final analysis above are based on the FI and are valid only for $N00N$ states.  In general, however, to assess the value of a state for metrology the QFI can be calculated and compared to the SIL.  For our state, $F_Q = 0.079$, giving a QCRB $\Delta\phi \ge 3.56/\sqrt{\nu}$.  For a classical state with the same average photon number as the reconstructed state, $\Delta\phi_\mathrm{SIL} = 1.81/\sqrt{\nu}$.  This shows again that, despite the implications of the post selected and renormalized analyses above, the classical limits to precision can never be surpassed with this state and arbitrary measurements.

Going beyond the classical performance limit with realistic quantum sensors requires stringent bounds on throughput, probe-state preparation and detection efficiencies. For heralded $N00N$ states, $f = \eta_\mathrm{p} (\eta\eta_\mathrm{d})^N$ must satisfy $f \ge \eta/ NV^{2}$. For the ideal two-photon situation ($V =1$) this implies $\eta_\mathrm{p}\eta\eta_\mathrm{d}^{2} \ge 1/2$.  This benchmark for heralded two-photon states provides a challenging goal that must be achieved to surpass the classical limit.  Even for perfect detectors ($\eta_\mathrm{d}=1$), this would require $\eta_\mathrm{p}\eta\ge1/2$.  Conversely, perfect preparation and transmission ($\eta_\mathrm{p}\eta=1$) implies that any detectors used must have $\eta_{d}\ge1/\sqrt{2}$.  To our knowledge, the above criteria have not yet been demonstrated together in a single experiment.

Quantum photonics using feasible laboratory technology has the potential to surpass the performance of classical techniques. Nonclassical behavior may be present in an experiment, but does not necessarily imply improved performance beyond classical limits. Here we have laid out an explicit approach to compare experimental results for precision metrology, which requires complete knowledge of the input state and number of experimental trials. We have demonstrated a scalable method to generate heralded entangled states of light for precision phase estimation, a key step to demonstrating a real-world improvement over the SQL. By fully characterizing this source we have calculated the ultimate precision that can be obtained and in doing so we have put bounds on the sensor parameters required for quantum enhanced phase estimation to be viable.

The authors acknowledge helpful discussions with N. Langford, R. Adamson, K. Banaszek, J. Lundeen.  This work has been supported under the EC integrated project Q-ESSENCE, US European Office of Aerospace Research (FA8655-09-1-3020), EPSRC (EP/H03031X/1), and the Royal Society.

\bibliography{realWorldQuantumSensors}

\begin{thebibliography}{36}%
\makeatletter
\providecommand \@ifxundefined [1]{%
 \@ifx{#1\undefined}
}%
\providecommand \@ifnum [1]{%
 \ifnum #1\expandafter \@firstoftwo
 \else \expandafter \@secondoftwo
 \fi
}%
\providecommand \@ifx [1]{%
 \ifx #1\expandafter \@firstoftwo
 \else \expandafter \@secondoftwo
 \fi
}%
\providecommand \natexlab [1]{#1}%
\providecommand \enquote  [1]{``#1''}%
\providecommand \bibnamefont  [1]{#1}%
\providecommand \bibfnamefont [1]{#1}%
\providecommand \citenamefont [1]{#1}%
\providecommand \href@noop [0]{\@secondoftwo}%
\providecommand \href [0]{\begingroup \@sanitize@url \@href}%
\providecommand \@href[1]{\@@startlink{#1}\@@href}%
\providecommand \@@href[1]{\endgroup#1\@@endlink}%
\providecommand \@sanitize@url [0]{\catcode `\\12\catcode `\$12\catcode
  `\&12\catcode `\#12\catcode `\^12\catcode `\_12\catcode `\%12\relax}%
\providecommand \@@startlink[1]{}%
\providecommand \@@endlink[0]{}%
\providecommand \url  [0]{\begingroup\@sanitize@url \@url }%
\providecommand \@url [1]{\endgroup\@href {#1}{\urlprefix }}%
\providecommand \urlprefix  [0]{URL }%
\providecommand \Eprint [0]{\href }%
\providecommand \doibase [0]{http://dx.doi.org/}%
\providecommand \selectlanguage [0]{\@gobble}%
\providecommand \bibinfo  [0]{\@secondoftwo}%
\providecommand \bibfield  [0]{\@secondoftwo}%
\providecommand \translation [1]{[#1]}%
\providecommand \BibitemOpen [0]{}%
\providecommand \bibitemStop [0]{}%
\providecommand \bibitemNoStop [0]{.\EOS\space}%
\providecommand \EOS [0]{\spacefactor3000\relax}%
\providecommand \BibitemShut  [1]{\csname bibitem#1\endcsname}%
\let\auto@bib@innerbib\@empty
\bibitem [{\citenamefont {Giovannetti}\ and\ \citenamefont
  {Lloyd}(2004)}]{glm04}%
  \BibitemOpen
  \bibfield  {author} {\bibinfo {author} {\bibfnamefont {V.}~\bibnamefont
  {Giovannetti}}\ and\ \bibinfo {author} {\bibfnamefont {S.}~\bibnamefont
  {Lloyd}},\ }\href@noop {} {\bibfield  {journal} {\bibinfo  {journal}
  {Science}\ }\textbf {\bibinfo {volume} {306}},\ \bibinfo {pages} {1330}
  (\bibinfo {year} {2004})}\BibitemShut {NoStop}%
\bibitem [{\citenamefont {Giovannetti}\ \emph {et~al.}(2006)\citenamefont
  {Giovannetti}, \citenamefont {Lloyd},\ and\ \citenamefont {Maccone}}]{glm06}%
  \BibitemOpen
  \bibfield  {author} {\bibinfo {author} {\bibfnamefont {V.}~\bibnamefont
  {Giovannetti}}, \bibinfo {author} {\bibfnamefont {S.}~\bibnamefont {Lloyd}},
  \ and\ \bibinfo {author} {\bibfnamefont {L.}~\bibnamefont {Maccone}},\
  }\href@noop {} {\bibfield  {journal} {\bibinfo  {journal} {Phys. Rev. Lett.}\
  }\textbf {\bibinfo {volume} {96}},\ \bibinfo {pages} {010401} (\bibinfo
  {year} {2006})}\BibitemShut {NoStop}%
\bibitem [{\citenamefont {Braunstein}\ and\ \citenamefont
  {Caves}(1994)}]{bc94}%
  \BibitemOpen
  \bibfield  {author} {\bibinfo {author} {\bibfnamefont {S.~L.}\ \bibnamefont
  {Braunstein}}\ and\ \bibinfo {author} {\bibfnamefont {C.~M.}\ \bibnamefont
  {Caves}},\ }\href@noop {} {\bibfield  {journal} {\bibinfo  {journal} {Phys.
  Rev. Lett.}\ }\textbf {\bibinfo {volume} {72}},\ \bibinfo {pages} {3439}
  (\bibinfo {year} {1994})}\BibitemShut {NoStop}%
\bibitem [{\citenamefont {Dorner~et al.}(2009)}]{dorner:09}%
  \BibitemOpen
  \bibfield  {author} {\bibinfo {author} {\bibfnamefont {U.}~\bibnamefont
  {Dorner~et al.}},\ }\href@noop {} {\bibfield  {journal} {\bibinfo  {journal}
  {Phys. Rev. Lett.}\ }\textbf {\bibinfo {volume} {102}},\ \bibinfo {pages}
  {040403} (\bibinfo {year} {2009})}\BibitemShut {NoStop}%
\bibitem [{\citenamefont {Datta~et al.}(2010)}]{datta:10}%
  \BibitemOpen
  \bibfield  {author} {\bibinfo {author} {\bibfnamefont {A.}~\bibnamefont
  {Datta~et al.}},\ }\href@noop {} {\bibfield  {journal} {\bibinfo  {journal}
  {arXiv:1012.0539v1 [quant-ph]}\ } (\bibinfo {year} {2010})}\BibitemShut
  {NoStop}%
\bibitem [{\citenamefont {Huelga~et al.}(1997)}]{Huelga97}%
  \BibitemOpen
  \bibfield  {author} {\bibinfo {author} {\bibfnamefont {S.~F.}\ \bibnamefont
  {Huelga~et al.}},\ }\href@noop {} {\bibfield  {journal} {\bibinfo  {journal}
  {Phys. Rev. Lett.}\ }\textbf {\bibinfo {volume} {79}},\ \bibinfo {pages}
  {3865} (\bibinfo {year} {1997})}\BibitemShut {NoStop}%
\bibitem [{\citenamefont {Shaji}\ and\ \citenamefont {Caves}(2007)}]{Shaji07}%
  \BibitemOpen
  \bibfield  {author} {\bibinfo {author} {\bibfnamefont {A.}~\bibnamefont
  {Shaji}}\ and\ \bibinfo {author} {\bibfnamefont {C.~M.}\ \bibnamefont
  {Caves}},\ }\href@noop {} {\bibfield  {journal} {\bibinfo  {journal} {Phys.
  Rev. A}\ }\textbf {\bibinfo {volume} {76}},\ \bibinfo {pages} {032111}
  (\bibinfo {year} {2007})}\BibitemShut {NoStop}%
\bibitem [{\citenamefont {Rubin}\ and\ \citenamefont
  {Kaushik}(2007)}]{Rubin07}%
  \BibitemOpen
  \bibfield  {author} {\bibinfo {author} {\bibfnamefont {A.~M.}\ \bibnamefont
  {Rubin}}\ and\ \bibinfo {author} {\bibfnamefont {S.}~\bibnamefont
  {Kaushik}},\ }\href@noop {} {\bibfield  {journal} {\bibinfo  {journal} {Phys.
  Rev. A}\ }\textbf {\bibinfo {volume} {75}},\ \bibinfo {pages} {053805}
  (\bibinfo {year} {2007})}\BibitemShut {NoStop}%
\bibitem [{\citenamefont {Huver}\ \emph {et~al.}(2008)\citenamefont {Huver},
  \citenamefont {Wildfeuer},\ and\ \citenamefont {Dowling}}]{Huver08}%
  \BibitemOpen
  \bibfield  {author} {\bibinfo {author} {\bibfnamefont {S.~D.}\ \bibnamefont
  {Huver}}, \bibinfo {author} {\bibfnamefont {C.~F.}\ \bibnamefont
  {Wildfeuer}}, \ and\ \bibinfo {author} {\bibfnamefont {J.~P.}\ \bibnamefont
  {Dowling}},\ }\href@noop {} {\bibfield  {journal} {\bibinfo  {journal} {Phys.
  Rev. A}\ }\textbf {\bibinfo {volume} {78}},\ \bibinfo {pages} {063828}
  (\bibinfo {year} {2008})}\BibitemShut {NoStop}%
\bibitem [{\citenamefont {Gilbert}\ \emph {et~al.}(2008)\citenamefont
  {Gilbert}, \citenamefont {Hamrick},\ and\ \citenamefont
  {Weinstein}}]{Gilbert08}%
  \BibitemOpen
  \bibfield  {author} {\bibinfo {author} {\bibfnamefont {G.}~\bibnamefont
  {Gilbert}}, \bibinfo {author} {\bibfnamefont {M.}~\bibnamefont {Hamrick}}, \
  and\ \bibinfo {author} {\bibfnamefont {Y.~S.}\ \bibnamefont {Weinstein}},\
  }\href@noop {} {\bibfield  {journal} {\bibinfo  {journal} {J. Opt. Soc. Am.
  B}\ }\textbf {\bibinfo {volume} {25}},\ \bibinfo {pages} {1336} (\bibinfo
  {year} {2008})}\BibitemShut {NoStop}%
\bibitem [{\citenamefont {Mitchell~et al.}(2004)}]{mls04}%
  \BibitemOpen
  \bibfield  {author} {\bibinfo {author} {\bibfnamefont {M.~W.}\ \bibnamefont
  {Mitchell~et al.}},\ }\href@noop {} {\bibfield  {journal} {\bibinfo
  {journal} {Nature}\ }\textbf {\bibinfo {volume} {429}},\ \bibinfo {pages}
  {161} (\bibinfo {year} {2004})}\BibitemShut {NoStop}%
\bibitem [{\citenamefont {Higgins~et al.}(2007)}]{higgins:07}%
  \BibitemOpen
  \bibfield  {author} {\bibinfo {author} {\bibfnamefont {B.~L.}\ \bibnamefont
  {Higgins~et al.}},\ }\href@noop {} {\bibfield  {journal} {\bibinfo  {journal}
  {Nature}\ }\textbf {\bibinfo {volume} {450}},\ \bibinfo {pages} {393}
  (\bibinfo {year} {2007})}\BibitemShut {NoStop}%
\bibitem [{\citenamefont {Eisenberg~et al.}(2005)}]{eisenberg:05}%
  \BibitemOpen
  \bibfield  {author} {\bibinfo {author} {\bibfnamefont {H.~S.}\ \bibnamefont
  {Eisenberg~et al.}},\ }\href@noop {} {\bibfield  {journal} {\bibinfo
  {journal} {Phys. Rev. Lett.}\ }\textbf {\bibinfo {volume} {94}},\ \bibinfo
  {pages} {090502} (\bibinfo {year} {2005})}\BibitemShut {NoStop}%
\bibitem [{\citenamefont {Resch~et al.}(2007{\natexlab{a}})}]{resch:07-1}%
  \BibitemOpen
  \bibfield  {author} {\bibinfo {author} {\bibfnamefont {K.~J.}\ \bibnamefont
  {Resch~et al.}},\ }\href@noop {} {\bibfield  {journal} {\bibinfo  {journal}
  {Phys. Rev. Lett.}\ }\textbf {\bibinfo {volume} {98}},\ \bibinfo {pages}
  {203602} (\bibinfo {year} {2007}{\natexlab{a}})}\BibitemShut {NoStop}%
\bibitem [{\citenamefont {Nagata~et al.}(2007)}]{nagata:07}%
  \BibitemOpen
  \bibfield  {author} {\bibinfo {author} {\bibfnamefont {T.}~\bibnamefont
  {Nagata~et al.}},\ }\href@noop {} {\bibfield  {journal} {\bibinfo  {journal}
  {Science}\ }\textbf {\bibinfo {volume} {316}},\ \bibinfo {pages} {726}
  (\bibinfo {year} {2007})}\BibitemShut {NoStop}%
\bibitem [{\citenamefont {Smith~et al.}(2008)}]{smith:08}%
  \BibitemOpen
  \bibfield  {author} {\bibinfo {author} {\bibfnamefont {B.~J.}\ \bibnamefont
  {Smith~et al.}},\ }in\ \href@noop {} {\emph {\bibinfo {booktitle}
  {CLEO/QELS}}},\ Vol.\ \bibinfo {volume} {1-9}\ (\bibinfo {year} {2008})\ p.\
  \bibinfo {pages} {3044}\BibitemShut {NoStop}%
\bibitem [{\citenamefont {Smith~et al.}(2009)}]{smith:09}%
  \BibitemOpen
  \bibfield  {author} {\bibinfo {author} {\bibfnamefont {B.~J.}\ \bibnamefont
  {Smith~et al.}},\ }\href@noop {} {\bibfield  {journal} {\bibinfo  {journal}
  {Optics Express}\ }\textbf {\bibinfo {volume} {17}},\ \bibinfo {pages}
  {13516} (\bibinfo {year} {2009})}\BibitemShut {NoStop}%
\bibitem [{\citenamefont {Matthews~et al.}(2009)}]{mpso09}%
  \BibitemOpen
  \bibfield  {author} {\bibinfo {author} {\bibfnamefont {J.~C.~F.}\
  \bibnamefont {Matthews~et al.}},\ }\href@noop {} {\bibfield  {journal}
  {\bibinfo  {journal} {Nature Photonics}\ }\textbf {\bibinfo {volume} {3}},\
  \bibinfo {pages} {346} (\bibinfo {year} {2009})}\BibitemShut {NoStop}%
\bibitem [{\citenamefont {Kim}\ \emph {et~al.}(2009)\citenamefont {Kim},
  \citenamefont {Park},\ and\ \citenamefont {Choi}}]{kim:09}%
  \BibitemOpen
  \bibfield  {author} {\bibinfo {author} {\bibfnamefont {H.}~\bibnamefont
  {Kim}}, \bibinfo {author} {\bibfnamefont {H.~S.}\ \bibnamefont {Park}}, \
  and\ \bibinfo {author} {\bibfnamefont {S.-K.}\ \bibnamefont {Choi}},\
  }\href@noop {} {\bibfield  {journal} {\bibinfo  {journal} {Optics Express}\
  }\textbf {\bibinfo {volume} {17}},\ \bibinfo {pages} {19720} (\bibinfo {year}
  {2009})}\BibitemShut {NoStop}%
\bibitem [{\citenamefont {Matthews~et al.}(2010)}]{matthews:10}%
  \BibitemOpen
  \bibfield  {author} {\bibinfo {author} {\bibfnamefont {J.~C.~F.}\
  \bibnamefont {Matthews~et al.}},\ }\href@noop {} {\bibfield  {journal}
  {\bibinfo  {journal} {arXiv:1005.5119v1 [quant-ph]}\ } (\bibinfo {year}
  {2010})}\BibitemShut {NoStop}%
\bibitem [{\citenamefont {Resch~et al.}(2007{\natexlab{b}})}]{resch:07}%
  \BibitemOpen
  \bibfield  {author} {\bibinfo {author} {\bibfnamefont {K.~J.}\ \bibnamefont
  {Resch~et al.}},\ }\href@noop {} {\bibfield  {journal} {\bibinfo  {journal}
  {Phys. Rev. Lett.}\ }\textbf {\bibinfo {volume} {98}},\ \bibinfo {pages}
  {223601} (\bibinfo {year} {2007}{\natexlab{b}})}\BibitemShut {NoStop}%
\bibitem [{\citenamefont {Afek}\ \emph {et~al.}(2010)\citenamefont {Afek},
  \citenamefont {Ambar},\ and\ \citenamefont {Silberberg}}]{afek:10-1}%
  \BibitemOpen
  \bibfield  {author} {\bibinfo {author} {\bibfnamefont {I.}~\bibnamefont
  {Afek}}, \bibinfo {author} {\bibfnamefont {O.}~\bibnamefont {Ambar}}, \ and\
  \bibinfo {author} {\bibfnamefont {Y.}~\bibnamefont {Silberberg}},\
  }\href@noop {} {\bibfield  {journal} {\bibinfo  {journal} {Science}\ }\textbf
  {\bibinfo {volume} {328}},\ \bibinfo {pages} {879} (\bibinfo {year}
  {2010})}\BibitemShut {NoStop}%
\bibitem [{\citenamefont {Sanders}\ and\ \citenamefont
  {Milburn}(1995)}]{sanders:95}%
  \BibitemOpen
  \bibfield  {author} {\bibinfo {author} {\bibfnamefont {B.~C.}\ \bibnamefont
  {Sanders}}\ and\ \bibinfo {author} {\bibfnamefont {G.~J.}\ \bibnamefont
  {Milburn}},\ }\href@noop {} {\bibfield  {journal} {\bibinfo  {journal} {Phys.
  Rev. Lett.}\ }\textbf {\bibinfo {volume} {75}},\ \bibinfo {pages} {2944}
  (\bibinfo {year} {1995})}\BibitemShut {NoStop}%
\bibitem [{\citenamefont {Boto~et al.}(2000)}]{boto:00}%
  \BibitemOpen
  \bibfield  {author} {\bibinfo {author} {\bibfnamefont {A.~N.}\ \bibnamefont
  {Boto~et al.}},\ }\href@noop {} {\bibfield  {journal} {\bibinfo  {journal}
  {Phys. Rev. Lett.}\ }\textbf {\bibinfo {volume} {85}},\ \bibinfo {pages}
  {2733} (\bibinfo {year} {2000})}\BibitemShut {NoStop}%
\bibitem [{\citenamefont {Bollinger~et al.}(1996)}]{bollinger:96}%
  \BibitemOpen
  \bibfield  {author} {\bibinfo {author} {\bibfnamefont {J.~J.}\ \bibnamefont
  {Bollinger~et al.}},\ }\href@noop {} {\bibfield  {journal} {\bibinfo
  {journal} {Phys. Rev. A}\ }\textbf {\bibinfo {volume} {54}},\ \bibinfo
  {pages} {R4649} (\bibinfo {year} {1996})}\BibitemShut {NoStop}%
\bibitem [{epa(ails)}]{epaps}%
  \BibitemOpen
  \href@noop {} {} (\bibinfo {year} {See appendices for details})\BibitemShut
  {NoStop}%
\bibitem [{\citenamefont {Holland}\ and\ \citenamefont
  {Burnett}(1993)}]{Holland:93}%
  \BibitemOpen
  \bibfield  {author} {\bibinfo {author} {\bibfnamefont {M.~J.}\ \bibnamefont
  {Holland}}\ and\ \bibinfo {author} {\bibfnamefont {K.}~\bibnamefont
  {Burnett}},\ }\href@noop {} {\bibfield  {journal} {\bibinfo  {journal} {Phys.
  Rev. Lett.}\ }\textbf {\bibinfo {volume} {71}},\ \bibinfo {pages} {1355}
  (\bibinfo {year} {1993})}\BibitemShut {NoStop}%
\bibitem [{\citenamefont {Dunningham}\ \emph {et~al.}(2002)\citenamefont
  {Dunningham}, \citenamefont {Burnett},\ and\ \citenamefont
  {Barnett}}]{dunningham:02}%
  \BibitemOpen
  \bibfield  {author} {\bibinfo {author} {\bibfnamefont {J.~A.}\ \bibnamefont
  {Dunningham}}, \bibinfo {author} {\bibfnamefont {K.}~\bibnamefont {Burnett}},
  \ and\ \bibinfo {author} {\bibfnamefont {S.~M.}\ \bibnamefont {Barnett}},\
  }\href@noop {} {\bibfield  {journal} {\bibinfo  {journal} {Phys. Rev. Lett.}\
  }\textbf {\bibinfo {volume} {89}},\ \bibinfo {pages} {150401} (\bibinfo
  {year} {2002})}\BibitemShut {NoStop}%
\bibitem [{\citenamefont {Mosley~et al.}(2008)}]{mosley:08}%
  \BibitemOpen
  \bibfield  {author} {\bibinfo {author} {\bibfnamefont {P.~J.}\ \bibnamefont
  {Mosley~et al.}},\ }\href@noop {} {\bibfield  {journal} {\bibinfo  {journal}
  {Phys. Rev. Lett.}\ }\textbf {\bibinfo {volume} {100}},\ \bibinfo {pages}
  {133601} (\bibinfo {year} {2008})}\BibitemShut {NoStop}%
\bibitem [{\citenamefont {Achilles~et al.}(2003)}]{achilles:03}%
  \BibitemOpen
  \bibfield  {author} {\bibinfo {author} {\bibfnamefont {D.}~\bibnamefont
  {Achilles~et al.}},\ }\href@noop {} {\bibfield  {journal} {\bibinfo
  {journal} {Opt. Lett.}\ }\textbf {\bibinfo {volume} {28}},\ \bibinfo {pages}
  {2387} (\bibinfo {year} {2003})}\BibitemShut {NoStop}%
\bibitem [{\citenamefont {Thomas-Peter}\ \emph {et~al.}(2009)\citenamefont
  {Thomas-Peter}, \citenamefont {Smith},\ and\ \citenamefont
  {Walmsley}}]{thomas-peter:09}%
  \BibitemOpen
  \bibfield  {author} {\bibinfo {author} {\bibfnamefont {N.~L.}\ \bibnamefont
  {Thomas-Peter}}, \bibinfo {author} {\bibfnamefont {B.~J.}\ \bibnamefont
  {Smith}}, \ and\ \bibinfo {author} {\bibfnamefont {I.~A.}\ \bibnamefont
  {Walmsley}},\ }in\ \href@noop {} {\emph {\bibinfo {booktitle} {CLEO/QELS}}},\
  Vol.\ \bibinfo {volume} {1-9}\ (\bibinfo {year} {2009})\ p.\ \bibinfo {pages}
  {2430}\BibitemShut {NoStop}%
\bibitem [{\citenamefont {Holevo}(1982)}]{HolevoAS1982psa}%
  \BibitemOpen
  \bibfield  {author} {\bibinfo {author} {\bibfnamefont {A.~S.}\ \bibnamefont
  {Holevo}},\ }\href@noop {} {\emph {\bibinfo {title} {Probabilistic and
  statistical aspects of quantum theory}}}\ (\bibinfo  {publisher}
  {North-Holland},\ \bibinfo {year} {1982})\BibitemShut {NoStop}%
\bibitem [{\citenamefont {Helstrom}(1976)}]{HelstromCW1976qde}%
  \BibitemOpen
  \bibfield  {author} {\bibinfo {author} {\bibfnamefont {C.~W.}\ \bibnamefont
  {Helstrom}},\ }\href@noop {} {\emph {\bibinfo {title} {Quantum detection and
  estimation theory}}}\ (\bibinfo  {publisher} {Academic Press},\ \bibinfo
  {year} {1976})\BibitemShut {NoStop}%
\bibitem [{\citenamefont {S.~L.~Braunstein}(1996)}]{BraunsteinSL1994gur}%
  \BibitemOpen
  \bibfield  {author} {\bibinfo {author} {\bibfnamefont {G.~J.~M.}\
  \bibnamefont {S.~L.~Braunstein}, \bibfnamefont {C.~M.~Caves}},\ }\href@noop
  {} {\bibfield  {journal} {\bibinfo  {journal} {Ann. Phys.}\ }\textbf
  {\bibinfo {volume} {247}},\ \bibinfo {pages} {135} (\bibinfo {year}
  {1996})}\BibitemShut {NoStop}%
\bibitem [{\citenamefont {Adamson~et al.}(2007)}]{asms07}%
  \BibitemOpen
  \bibfield  {author} {\bibinfo {author} {\bibfnamefont {R.~B.~A.}\
  \bibnamefont {Adamson~et al.}},\ }\href@noop {} {\bibfield  {journal}
  {\bibinfo  {journal} {Phys. Rev. Lett.}\ }\textbf {\bibinfo {volume} {98}},\
  \bibinfo {pages} {043601} (\bibinfo {year} {2007})}\BibitemShut {NoStop}%
\bibitem [{\citenamefont {Banaszek~et al.}(1999)}]{bdps99}%
  \BibitemOpen
  \bibfield  {author} {\bibinfo {author} {\bibfnamefont {K.}~\bibnamefont
  {Banaszek~et al.}},\ }\href@noop {} {\bibfield  {journal} {\bibinfo
  {journal} {Phys. Rev. A}\ }\textbf {\bibinfo {volume} {61}},\ \bibinfo
  {pages} {010304} (\bibinfo {year} {1999})}\BibitemShut {NoStop}%
\end{thebibliography}%

\clearpage

\appendix

\section{Fisher information and phase uncertainty}

The amount of information about the phase $\phi$ between two modes in an interferometer that can be extracted by a given experimental configuration is quantified by the Fisher information
\begin{equation}
F(\phi) = \sum_j \frac{1}{p_j ( \phi )} \left|\frac{\partial p_j (\phi)}{\partial \phi}\right|^{2}.
\label{eq:1a}
\end{equation}
Here the probabilities $p_j(\phi)$ correspond to a particular outcome
$j$ of a measurement described by a set of operators $\{ \hat\Pi_j \}$
when the actual parameter to be estimated has a value $\phi $. The
precision with which $\phi$ can be estimated is determined by the
Cram\'{e}r-Rao bound (CRB)~\cite{HolevoAS1982psa},
\begin{equation}
\Delta \phi \ge 1 / \sqrt{\nu F(\phi)},
\label{crb}
\end{equation}
which can be achieved for a large number of trials $\nu$ and an
unbaised estimator, such as a maximum likelihood method. The CRB is
bounded from below by the quantum Cram\'{e}r-Rao bound (QCRB), which
is independent of the measurement and depends only on the state used,
\begin{equation}
\Delta \phi \ge 1 / \sqrt{\nu F(\phi)} \ge 1 / \sqrt{\nu F_Q(\phi)},
\label{qcrb}
\end{equation}
where $F_Q\left(\phi \right)=F_Q\left(\hat{\rho}(\phi)\right)$ is the
quantum Fisher information~\cite{HelstromCW1976qde, HolevoAS1982psa, BraunsteinSL1994gur, bc94}. Note that in general both the FI and the QFI depend on the value of the  parameter $\phi$. Only in certain special cases are they independent of the parameter.

In a perfect experiment, an $N$-fold detection event corresponds to projection onto an $N$-photon $N00N$ state, $\ket{\Psi_{\pm}} = (|N,0\rangle \pm |0,N\rangle ) / \sqrt{2}$, where $\ket{m,n}$ denotes $m$ ($n$) photons in mode $a$ ($b$). This measurement set has three possible outcomes: $j=\pm$ (detection of $\ket{\Psi_\pm}$) and $j=0$ (otherwise).  The probabilities associated with these measurement outcomes for a $N00N$ state input are $p_{\pm}(\phi) = [1\pm \cos(N\phi)]/2$. In practice, it is difficult to realize single-mode sensors and this can lead to degradation of interference due to unmeasured distinguishing information of the input state.  Measurement outcomes therefore typically have probabilities
\begin{equation}
p_{\pm}(\phi) = f[1\pm V \cos(N\phi)]/2,
\end{equation}
and $p_{0}(\phi) = 1-p_{+}(\phi) - p_{-}(\phi)$, where $f$ and $V$ both depend upon the presence of distinguishing information in the input state. Here $f$ is related to the probability that an input state leads to an $N$-fold detection event, and $V$ is the fringe visibility. The Fisher information for this
configuration is given by
\begin{equation}
F(\phi) = f N^2 V^2 \frac{ \sin^2(N \phi) } { 1- V^2 \cos^2 (N\phi) },
\label{eq:2}
\end{equation}
where we have made use of the first derivatives of $p_{\pm}(\phi)$
with respect to $\phi$,
\begin{equation}
\frac{\partial p_{\pm} (\phi)}{\partial \phi} = \mp \frac{ f } {2} N V \sin(N \phi).
\label{eq:3}
\end{equation}
Note that the denominator in Eq. (\ref{eq:2}) is greater than or equal
to $\sin^2(N \phi)$ for $0\le V \le 1$, which sets a bound on the
Fisher information,
\begin{equation}
F(\phi) \le f V^2 N^2.
\label{eq:4}
\end{equation}
The phase uncertainty that can be achieved with $\nu$ trials is then
bounded by the Cram\'{e}r-Rao bound
\begin{equation}
\Delta \phi \ge 1/ (\sqrt{\nu f} N V).
\label{eq:5}
\end{equation}
Comparison with the precision that can be achieved in a classical
approach with $N$ particles, i.e. the standard quantum limit (SQL),
\begin{equation}
\Delta \phi_{\mathrm{SQL}} = 1/ \sqrt{\nu N},
\label{eq:5b}
\end{equation}
implies that in order to demonstrate precision beyond the SQL,
\begin{equation}
V \ge 1 / \sqrt{f N}.
\label{eq:6}
\end{equation}

\section{Effects of loss and detection inefficiencies on measurement precision}

We consider the case of a single spectral-mode sensor, since this enables $f$ to be calculated straightforwardly and allows comparison with previous work involving super-resolution and super-sensitivity criteria. For the case of no loss and perfect detectors, we define $f = \eta_\mathrm{p}$, the production efficiency of the single-mode $N00N$ state. When there is balanced loss in the interferometer, characterized by the transmissivity $\eta$ and detector efficiency $\eta_\mathrm{d}$, the probability that an $N$-photon $N00N$ state is successfully detected is decreased by a factor $(\eta\eta_\mathrm{d})^N$, giving $f = \eta_p (\eta\eta_\mathrm{d})^N$. In the classical case, the probability that a coherent state is transmitted is simply $\eta$ so that the classical phase uncertainty is bounded by $\Delta\phi_\mathrm{SIL} = 1 / \sqrt{\nu \eta N}$~\cite{dorner:09}.  This bound is the best precision that can be achieved by using the same channel with  a classical input state and perfect detectors.

$f$ can be estimated from the ratio of successful $N00N$ detection events to herald events, giving $\eta\eta_\mathrm{d} = \sqrt[N]{f/\eta_p}$. To surpass the performance allowed classically, i.e. $\Delta \phi_{\mathrm{SIL}} \ge 1/(\sqrt{\nu f} VN)$, the fringe visibility must be
\begin{equation}
V \ge \sqrt{\frac{\eta}{f N}}.
\label{eq:8}
\end{equation}

\section{Time required to achieve equal precision: post-selected vs. non-post-selected}
\label{app:time}

In this section we calculate the number of experimental runs, which is equivalent to the time required to perform the experiment, which are necessary for a post-selected $N00N$ state scheme to achieve the same level of precision as a non-post-selected scheme when losses and imperfect detectors (characterized by $\eta$ and $\eta_\mathrm{d}$) and other imperfections (characterized by $V$ and $\eta_p$) are present. The apparent precision of a post-selected  $N00N$ state scheme is given by
\begin{equation}
\Delta \phi_\mathrm{ps} = \frac{1}{\sqrt{\nu'} V N},
\label{eq:prec_post}
\end{equation}
where $N$ is the number of photons detected in coincidence events at the {\em output} of the interferometer. Instead, if we consider the number of photons $N$ sent into the interferometer as the relevant resource we obtain
\begin{equation}
\Delta \phi_\mathrm{min} = \frac{1}{\sqrt{\nu f} V N}.
\label{eq:prec_non_post}
\end{equation}
By comparing Eqs.~(\ref{eq:prec_post}) and (\ref{eq:prec_non_post}) it follows that we need 
\begin{equation}
\nu' = \frac{\nu}{f} = \frac{\nu}{\eta_p(\eta\eta_\mathrm{d})^N}
\end{equation}
more experimental trials in the post-selected scheme to obtain precision equal to the lossless case, which scales exponentially with the size of the $N00N$-state. 

Similarly we can compare a $N00N$-state scheme with a classical scheme of precision
\begin{equation}
\Delta \phi_\mathrm{SIL} = \frac{1}{\sqrt{\nu \eta N}}.
\label{eq:prec_class}
\end{equation}
Comparing this with Eq.~(\ref{eq:prec_non_post}) we see that the $N00N$-state scheme requires
\begin{equation}
\frac{\nu\eta}{\eta_p(\eta\eta_\mathrm{d})^N V^2 N}
\end{equation}
trials to achieve the classical precision. From this point of view the $N00N$-state scheme beats the classical scheme only if $\eta/(\eta_p(\eta\eta_\mathrm{d})^NV^2N)<1$, which is challenging to achieve in practice. 

\section{Heralded Holland-Burnett state generation}

Holland-Burnett or twin-Fock states \cite{Holland:93}, which we denote HB($N=2K$), where $K\in \{1,2,3,\ldots\}$, can be prepared by sending $K$ photons to both inputs of a 50:50 beam splitter. The state of the field at the two output modes of the beam splitter is given by
\begin{equation}
| \Psi _{K} \rangle = \sum_{m=0}^{K} A_{m,K} |2m\rangle_{a} |2(K - m)\rangle_{b},
\label{HB}
\end{equation}
where the coefficients are
$$A_{m,K} = (-1)^{K-m}\left[ \binom{2m}{m} \binom{2(K-m)}{K-m} \left( \frac{1}{2}\right)^{2K} \right]^{1/2}.$$
This notation differs slightly from the standard notation, where $K = N$, and there are a total of $N_{\rm{tot}}=2N$ photons instead of $N$. We choose our notation to clarify the comparison between $N00N$ states and the HB states. Note that the HB state has similar structure to the states examined for optimal phase estimation in lossy interferometers in Ref. \cite{dorner:09}.

The heralded HB-state source is based upon generation of heralded Fock
states by means of parametric downconversion in Potassium Dihydrogen
Phosphate (KDP) \cite{mosley:08}. A crucial feature of this nonlinear
source is that the modes are entangled only in photon number and have
no entanglement in any other degrees of freedom. This allows detection
of $N$ photons in one mode to herald, in principle, $N$ photons in the
other mode in a pure spatiotemporal quantum state. No spectral
filtering is needed to increase the state purity of the heralded
photons. This scheme is therefore scalable, since there is no decrease
in heralding efficiency arising from discarding photons by filtration.
The use of high-efficiency number-resolving heralding detectors
enables production of precisely the target HB state abrogating the
need for ancillary assumptions about contamination from higher or
lower photon numbers.

A mode-locked Ti:Sapphire laser system operating at 80 MHz provides
100 fs duration pulses centered at a wavelength of 830 nm. These
pulses are frequency doubled in Beta Barium Borate (BBO) to a
wavelength of 415 nm. After filtering any residual infrared radiation,
the UV-pulses are used to concurrently pump two KDP crystals oriented
for type-II collinear parametric downconversion (PDC)
\cite{mosley:08}. The orthogonally polarized PDC modes are split using
polarizing beam splitters. The vertically polarized mode of each PDC
source is detected by a fiber-coupled avalanche photodiode (APD)
serving as a herald signal for the horizontal modes. After rotating
one of the horizontal modes to vertical polarization, the heralded
modes are sent to a fiber-coupled polarizing beam splitter (FPBS)
where they are combined into a single spatial mode. Each heralded beam
and the FPBS output are passed through polarization compensation wave
plates to ensure appropriate polarization after passing through the
optical fibers. The FPBS output has additional polarization
compensation. Successful preparation of the heralded state is signaled
by a coincidence detection event between these two heralding
detectors. When only one photon is registered at each herald detector,
this heralds the state $\ket{1,1}_\mathrm{HV})$ where
$\ket{n,m}_\mathrm{HV}$ denotes $n$ ($m$) horizontally (vertically)
 polarized photons. This is equivalent to the two-photon HB state
$(\ket{2,0}_{+-} - \ket{0,2}_{+-})/\sqrt{2}$, where $+,-$ denotes the
$45^{\circ}$ rotated basis. Note that single photons are typically
heralded with an efficiency of 0.18 in this experimental
configuration.

\section{Interferometry}\label{phase}
We used a polarization-based interferometer to apply phase shifts
between $+45^{\circ}$ and $-45^{\circ}$ plane polarization modes. This
consists of a half-wave plate situated between two quarter-wave
plates. The quarter-wave plates were initially aligned to rotate $+$
into $-$ without the half-wave plate. The addition of the half-wave
plate allows application of the phase, $\phi$, being twice the angle
of the half-wave plate. The $+$ and $-$ modes are then interfered on a
polarizing beam splitter for analysis.

\section{State tomography of the heralded state}\label{tomo}
We characterize the polarization and photon number degrees of freedom
of the heralded state by performing state tomography \cite{asms07,
 thomas-peter:09}. A half- (HWP) is followed by a quarter-wave plate
(QWP), the angles of which determine the measurement implemented. The
HWP and QWP are followed by a polarizing beam splitter (PBS) with the
transmitted mode coupled into a single mode fiber-coupled APD and the
reflected mode coupled into a single-mode fiber beam splitter (50:50)
with outputs connected to fiber-coupled APDs. This allows partial
photon number resolution, giving information about the different
photon number subspaces when measured in coincidence with the herald
signal. The wave plate settings used, $\{\theta, \phi\}$, were
$\{0^{\circ},0^{\circ}\}$, $\{0^{\circ},11.25^{\circ}\}$, $\{0^{\circ},22.5^{\circ}\}$, $\{0^{\circ},45^{\circ}\}$, $\{22.5^{\circ},0^{\circ}\}$,
$\{22.5^{\circ},22.5^{\circ}\}$, $\{22.5^{\circ},45^{\circ}\}$, and $\{45^{\circ},22.5^{\circ}\}$, where $\theta$ and
$\phi$ are the angles of the QWP and HWP respectively. This forms an
over-complete set of measurements on the whole space. The state was
reconstructed by maximum likelihood estimation \cite{bdps99} using the
measurement outcomes. In order to check the consistency of the
reconstructed density matrix, we calculate an overlap of $0.85$ of the
two-photon subspace with the ideal two-photon HB state,
Fig.~\ref{fig:rho}c, in agreement with the Hong-Ou-Mandel interference
visibility.

\section{Quantum Fisher information of the reconstructed state}

For the reconstructed density matrix in our experiment, the quantum
Fisher information is given by~\cite{BraunsteinSL1994gur}
\begin{equation}
F_Q = 2 \sum_{k,l} \frac{(p_k - p_l)^2}{p_k + p_l} |\langle \xi_k | \hat{n} | \xi_l \rangle |^2,
\label{qfi}
\end{equation}
where $\hat{n}$ is the photon number operator in the sensor arm of the
interferometer, and $\left\{ p_k \right\}$ and $\left\{ | \xi_k
 \rangle \right\}$ are the eigenvalues and corresponding eigenvectors
of the reconstructed density matrix.

\end{document}